\documentclass[conference]{IEEEtran}

\usepackage{graphicx}
\usepackage{booktabs}
\usepackage{xurl}
\usepackage{algorithm}
\usepackage{algorithmic}
\usepackage{amsmath}
\usepackage{eso-pic}

\usepackage{caption}
\makeatletter
\newcommand\notsotiny{\@setfontsize\notsotiny\@vipt\@viipt}
\makeatother

\AddToShipoutPictureBG*{
  \AtPageUpperLeft{
    \raisebox{-1.2cm}{\hspace{2cm}\fbox{\parbox{0.95\textwidth}{\small This paper was originally presented at IEEE ICC 2026. An extended version of this work has been published in IEEE Access.\\
    For citation, please refer to the IEEE Access version: D.~Chiba, H.~Nakano, and T.~Koide, ``PhishLumos: From a Single URL to Campaign-Level Phishing Mitigation,'' IEEE Access, 2026, https://doi.org/10.1109/ACCESS.2026.3696597.}}}
  }
  \AtPageLowerLeft{
    \raisebox{1.2cm}{\hspace{2cm}\parbox{0.95\textwidth}{\footnotesize © 2026 IEEE. Personal use of this material is permitted. Permission from IEEE must be obtained for all other uses, in any current or future media, including reprinting/republishing this material for advertising or promotional purposes, creating new collective works, for resale or redistribution to servers or lists, or reuse of any copyrighted component of this work in other works.}}
  }
}

\begin{document}

\title{PhishLumos: An Adaptive Multi-Agent System for Proactive Phishing Campaign Mitigation}

\author{
\IEEEauthorblockN{Daiki Chiba\IEEEauthorrefmark{1},
Hiroki Nakano\IEEEauthorrefmark{2}, and
Takashi Koide\IEEEauthorrefmark{2}}
\IEEEauthorblockA{\IEEEauthorrefmark{1}Tokyo Metropolitan University, Tokyo, Japan\\
Email: daiki.chiba@ieee.org}
\IEEEauthorblockA{\IEEEauthorrefmark{2}NTT Security Holdings Corporation \& NTT, Inc., Tokyo, Japan}
}

\maketitle

\begin{abstract}
Phishing attacks are a significant societal threat, disproportionately harming vulnerable populations and eroding trust in essential digital services. Current defenses are often reactive, failing against modern evasive tactics like cloaking that conceal malicious content. To address this, we introduce PhishLumos, an adaptive multi-agent system that proactively mitigates entire attack campaigns. It confronts a core cybersecurity imbalance: attackers can easily scale operations, while defense remains an intensive expert task. Instead of being blocked by evasion, PhishLumos treats it as a critical signal to investigate the underlying infrastructure. Its Large Language Model (LLM)-powered agents uncover shared hosting, certificates, and domain registration patterns. On real-world data, our system identified 100\% of campaigns in the median case, over a week before their confirmation by cybersecurity experts. PhishLumos demonstrates a practical shift from reactive URL blocking to proactive campaign mitigation, protecting users before they are harmed and making the digital world safer for all.
\end{abstract}

\section{Introduction}
Phishing attacks, which use social engineering to steal confidential data, represent a persistent and adaptable security threat.
The impact is substantial: total losses from cybercrime in the U.S. exceeded \$12.5 billion in 2023 alone.
Within this landscape, phishing stands out as the most frequently reported type of crime, highlighting its central role as a societal threat~\cite{fbi2023report1_ic3}.
These attacks deepen the digital divide by disproportionately harming vulnerable populations, including older adults and those with less technical expertise~\cite{fbi2023report2_elder}.
Critically, this erosion of trust in essential digital services---from banking to healthcare---threatens not just economic stability but also digital equity in our increasingly online society.
Defensive strategies have long relied on tools like blocklists, which are typically populated based on data from automated security scanners.
However, today's attackers employ sophisticated evasive tactics, such as cloaking (delivering different content to scanners versus users), which makes these established methods less effective.
This situation demands more advanced, proactive, and resilient solutions to counter these modern threats.

The limitations of standard detection are particularly clear with content-based techniques, which become ineffective when webpages offer little inspectable data.
Attackers exploit this weakness through evasive strategies that minimize analyzable content, from hiding pages behind CAPTCHAs to returning benign \texttt{404} errors to scanners~\cite{DBLP:conf/uss/Teoh0LHD24,DBLP:conf/sp/ZhangOCSJWSKBWS21,DBLP:conf/sp/OestSDAWT19,DBLP:conf/uss/AcharyaV21}.
The widespread availability of phishing kits intensifies this struggle, ensuring defenses remain reactive.
This dependency on content is a core issue that persists even with modern approaches.
The application of powerful Large Language Models (LLMs) to phishing detection, for example, often presumes the availability of rich content for analysis~\cite{DBLP:conf/uss/Liu0TLHD24}.
Consequently, their performance degrades in \textit{content-inaccessible} scenarios.
When faced with a cloaked page, an LLM cannot function as intended because there is no meaningful content from which to reason about the page's purpose or identify targeted brands.
This reveals a critical need for new methods that can uncover entire malicious campaigns from incomplete or deliberately obscured data.

\begin{figure}[t]
    \centering
    \includegraphics[width=1.0\columnwidth]{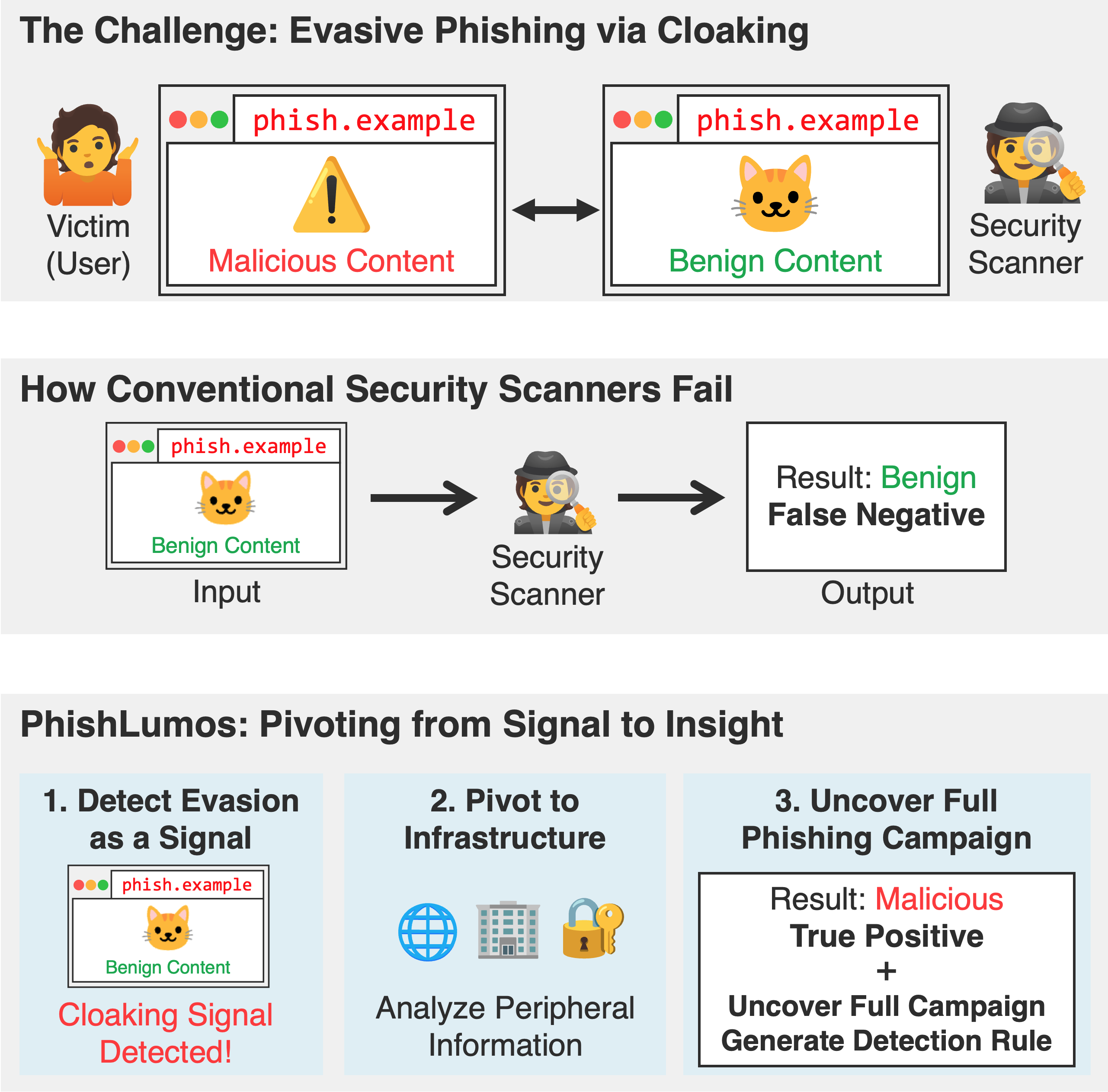}
    \caption{The core concept of PhishLumos.}
    \label{fig1}
\end{figure}

We introduce PhishLumos, a multi-agent system that confronts a core cybersecurity asymmetry: easy attack scaling versus resource-heavy defense.
Attackers often deploy coordinated campaigns---launching numerous similar phishing sites simultaneously to maximize impact.
PhishLumos automates the expert-level investigation of these entire campaigns from a single URL.
As shown in Figure~\ref{fig1}, our system treats evasion not as an obstacle but as a key signal to investigate the attacker's infrastructure.
This pivot allows it to uncover campaigns from minimal clues using LLM-powered agents that verify threats, characterize campaigns, and generate validated detection rules.
Our contribution is providing actionable intelligence for campaign mitigation.
This work represents a shift towards proactive defense, aiming to protect vulnerable users by mitigating campaigns before they cause significant harm.

\section{Related Work}
Detecting coordinated phishing campaigns is difficult when evasion hides page content from automated scanners. Prior work spans (1) content-based detectors, (2) measurements and countermeasures for cloaking, and (3) LLM and multimodal methods. Most assume accessible content and target single URLs, leaving little support for campaign discovery from sparse signals. PhishLumos addresses this gap by starting from minimal evidence and producing campaign-level rules.

\noindent\textbf{Content-Based Phishing Detection.}
Visual similarity, logo recognition, and intent inference drive recent detectors such as VisualPhishNet~\cite{DBLP:conf/ccs/AbdelnabiKF20}, Phishpedia~\cite{DBLP:conf/uss/LinLDNCLSZD21}, and PhishIntention~\cite{DBLP:conf/uss/Liu0YNDD22}. These systems are effective when pages expose text, layout, or images, but performance degrades when content is hidden or unavailable. Unlike these methods, PhishLumos operates in content-inaccessible cases (scanners cannot retrieve meaningful content) by pivoting to infrastructure signals and grouping pages into campaigns.

\noindent\textbf{Evasion Techniques and Countermeasures.}
Work on cloaking and client-side evasions includes CrawlPhish~\cite{DBLP:conf/sp/ZhangOCSJWSKBWS21} and Invernizzi \textit{et~al.}~\cite{DBLP:conf/sp/InvernizziTKCPB16}; controlled measurements in PhishFarm~\cite{DBLP:conf/sp/OestSDAWT19} and PhishTime~\cite{DBLP:conf/uss/OestSZWTSD20} showed their effectiveness. Tools such as PhishDecloaker~\cite{DBLP:conf/uss/Teoh0LHD24}, PhishPrint~\cite{DBLP:conf/uss/AcharyaV21}, and PhishParrot~\cite{phishparrot} aim to bypass or analyze these tactics. PhishLumos differs by treating inaccessibility as a trigger to investigate shared hosting, certificates, and registration patterns, then emitting rules that describe the campaign rather than the page.

\noindent\textbf{Multimodal Approaches and LLM Utilization.}
LLM and multimodal systems use crawled page content to aid detection or brand attribution, e.g., PhishLLM~\cite{DBLP:conf/uss/Liu0TLHD24}, KnowPhish~\cite{DBLP:conf/uss/LiHDLCOLH24}, ScamFerret~\cite{DBLP:conf/dimva/NakanoKC25}, ChatSpamDetector~\cite{chatspamdetector}, and ChatPhishDetector~\cite{DBLP:journals/access/KoideNC24}. These approaches work best when HTML, text, or screenshots are available. In contrast, PhishLumos tasks LLM agents to query external intelligence, correlate infrastructure clues across URLs, and validate candidate rules, enabling campaign-level detection even when page content is unavailable.

\section{Proposed System}

\subsection{System Overview}
\begin{figure}[t]
    \centering
    \includegraphics[width=1.0\columnwidth]{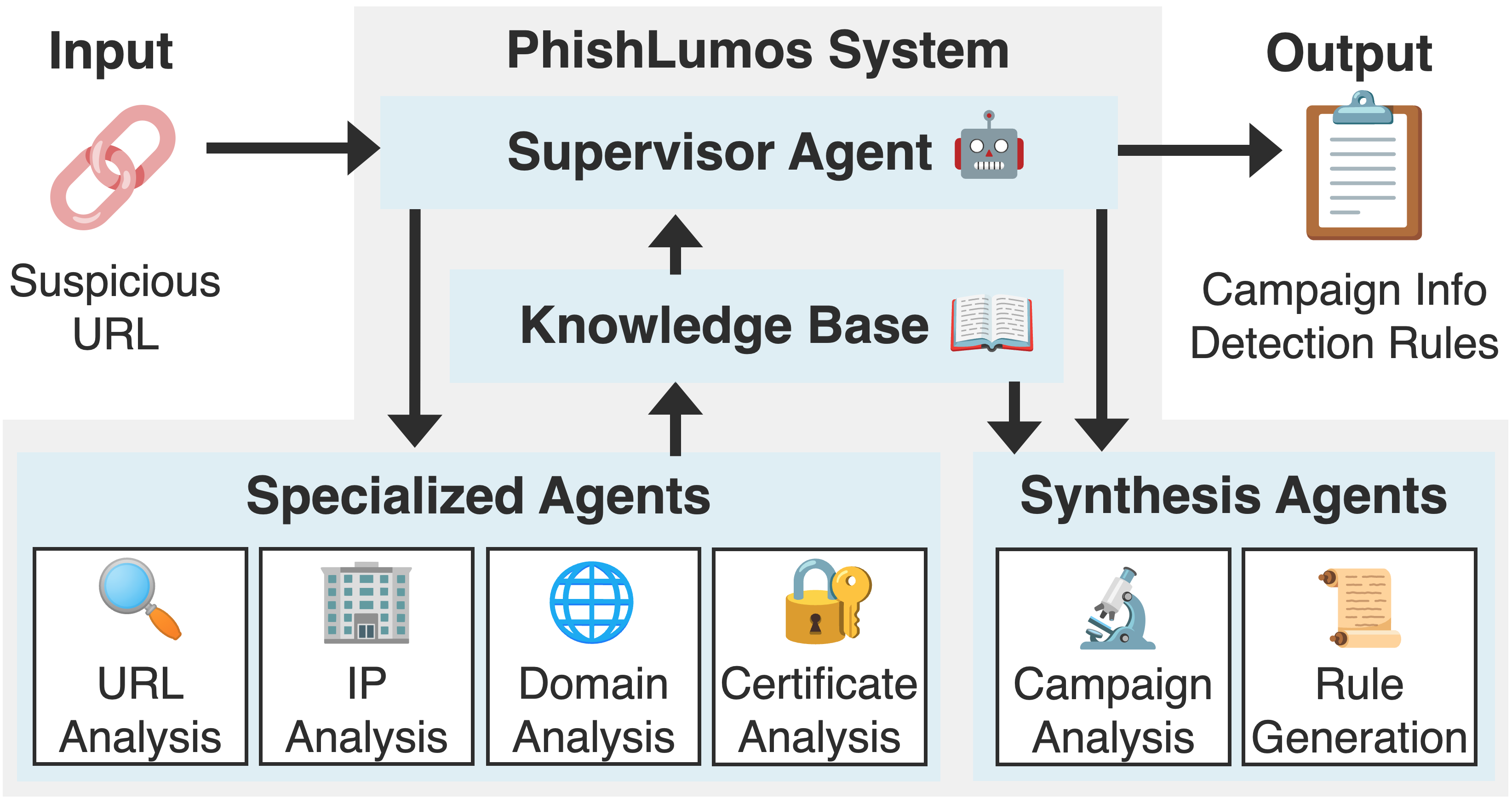}
    \caption{System Architecture of PhishLumos.}
    \label{fig2}
\end{figure}

We introduce PhishLumos, an adaptive multi-agent system for detecting advanced phishing attacks, particularly those employing evasive techniques that limit content availability.
Figure~\ref{fig2} illustrates the system architecture.
Unlike traditional methods that struggle with insufficient data, PhishLumos is designed for such content-inaccessible scenarios.
It performs what we term \textit{Pivotal Analysis}: an adaptive process where LLM-powered agents shift the investigative focus from an evasive URL to its underlying infrastructure footprints.
This involves targeted analysis of various external datasets, including historical Domain Name System (DNS) records, Internet Protocol (IP) address information, and Transport Layer Security (TLS) certificate data.

The architecture is coordinated by a central \textit{Supervisor Agent}, which directs a team of agents broadly categorized into two roles.
\textit{Specialized Agents} focus on data acquisition from distinct investigative areas (URL, domain, IP, and certificate analysis), while \textit{Synthesis Agents} are responsible for higher-level reasoning, such as profiling the campaign and generating detection rules.
The process begins when PhishLumos receives a URL.
The Supervisor's LLM evaluates the initial data to identify the most promising pivot points for further investigation.
For instance, if an initial scan reveals an IP address previously associated with malicious domains, the Supervisor prioritizes tasking the IP Analysis Agent.
Each agent employs an LLM to process information from its designated tools and returns structured findings.
These findings are continuously aggregated into a central \textit{Knowledge Base}---a dynamic, in-session repository of all collected evidence---allowing the Supervisor to build a comprehensive view of the attack.
Based on the evolving Knowledge Base, the Supervisor dynamically sequences subsequent tasks to systematically uncover the campaign's footprint.

\begin{table}[t]
    \centering
    \notsotiny
    \caption{PhishLumos agent roles and responsibilities.}
    \label{tab:agent_roles}
    {\renewcommand\arraystretch{0.75}
    \begin{tabular}{ll}
    \toprule
    \textbf{Agent} & \textbf{Goal / Input / Output} \\
    \midrule
    Supervisor & Goal: Orchestrate overall analysis flow. Input: URL. Output: Final Report.\\
    \midrule
    URL Analysis & Goal: Get historical scan data for a URL.\\
                 & Input: URL. Output: Scan history, status, IP addresses, brands. \\
    \midrule
    Domain Analysis & Goal: Analyze historical DNS/IP data.\\
                    & Input: Domain Name. Output: IP address history, stability patterns. \\
    \midrule
    IP Analysis & Goal: Analyze domains/activity on an IP.\\
                & Input: IP Address. Output: Associated domains, past malicious activity. \\
    \midrule
    Certificate Analysis & Goal: Analyze TLS certificate history.\\
                         & Input: Domain Name. Output: Issuer patterns, shared certificates. \\
    \midrule
    Campaign Analysis & Goal: Synthesize all findings to profile a campaign.\\
    & Input: All collected data. Output: Campaign type, key indicators. \\
    \midrule
    Rule Generation & Goal: Generate and validate detection rules.\\
    & Input: Key indicators. Output: Validated search queries. \\
    \bottomrule
    \end{tabular}
    }
\end{table}

\subsection{Threat Model}
Our threat model targets large-scale phishing campaigns designed to steal sensitive user credentials, which pose a significant socio-technical challenge.

\noindent\textbf{Social Dimension.}
These attacks erode digital equity by targeting vulnerable populations, such as older adults.
Beyond financial loss, they shatter the trust required to use essential online services, risking social isolation and loss of autonomy~\cite{fbi2023report2_elder}.

\noindent\textbf{Technical Dimension.}
We assume attackers use widely available phishing kits.
A key feature of these kits is sophisticated cloaking logic, which delivers malicious content to end-users while presenting benign content to automated scanners.
We consider email and Short Message Service (SMS) as primary distribution channels.
The analytical focus of PhishLumos is on characterizing and mitigating these coordinated campaigns that deploy many similar, cloaked sites concurrently, rather than isolated, one-off attacks.

\subsection{Core Components and Workflow}
The system's architecture is founded on two core principles: the integration of LLMs for advanced reasoning within each agent, and a structured collaboration framework enabling a flexible, state-aware workflow.
This design allows PhishLumos to adapt its investigative strategy in response to emerging findings.

The workflow begins with initial reconnaissance by the URL Analysis Agent, as directed by the Supervisor.
The Supervisor's LLM then evaluates these preliminary results to determine the most effective subsequent actions.
This involves dynamically assigning tasks to a team of agents, categorized as Specialized Agents for data gathering and Synthesis Agents for higher-level analysis.
As detailed in Table~\ref{tab:agent_roles}, each agent has a specific function and uses dedicated tools to query threat intelligence sources.
The findings from each agent update the central Knowledge Base.
This process enables PhishLumos to reason over a highly specific and contextual evidence graph, building a comprehensive threat profile from what may initially appear to be fragmented data.

\subsection{Adaptive Pivotal Analysis}

\begin{algorithm}[t]
    \notsotiny
    \caption{PhishLumos Adaptive Orchestration Logic}
    \label{alg:supervisor}
    \begin{algorithmic}[1]
    \STATE \textbf{Input:} Initial URL $u_0$
    \STATE \textbf{Initialize:} Knowledge Base $KB \leftarrow \emptyset$, History $H \leftarrow \emptyset$
    \STATE \COMMENT{Initial data acquisition}
    \STATE $task \leftarrow (\text{URLAnalysisAgent}, u_0)$
    \STATE $findings \leftarrow \text{ExecuteTask}(task)$
    \STATE $KB \leftarrow KB \cup findings$
    \STATE $H \leftarrow H \cup \{(task, findings)\}$
    \STATE \COMMENT{Iterative knowledge-driven analysis loop}
    \WHILE{$\text{InvestigationIsActive}(KB, H)$}
        \STATE \COMMENT{Supervisor assesses the comprehensive state}
        \STATE $action, params \leftarrow \text{LLM.DecideNextAction}(KB, H)$
        \STATE \COMMENT{e.g., assess cloaking or suspicious infra patterns}
        
        \IF{$action = \text{PIVOT\_TO\_INFRASTRUCTURE}$}
            \STATE $agent, \text{pivot\_data} \leftarrow params$
            \STATE $task \leftarrow (agent, \text{pivot\_data})$
            \STATE $findings \leftarrow \text{ExecuteTask}(task)$
            \STATE $KB \leftarrow KB \cup findings$
            \STATE $H \leftarrow H \cup \{(task, findings)\}$
        \ELSIF{$action = \text{CAMPAIGN\_ANALYSIS}$}
            \STATE \textbf{break} \COMMENT{Exit loop to synthesize}
        \ELSE
            \STATE \textbf{break} \COMMENT{Terminate if inconclusive}
        \ENDIF
    \ENDWHILE
    
    \STATE \COMMENT{Final synthesis and rule generation}
    \STATE $\text{campaign\_profile} \leftarrow \text{CampaignAnalysisAgent}(KB)$
    \IF{$\text{IsCampaignIdentified}(\text{campaign\_profile})$}
        \STATE $rules \leftarrow \text{RuleGenerationAgent}(\text{campaign\_profile})$
        \STATE \textbf{return} $\text{campaign\_profile}, rules$
    \ELSE
        \STATE \textbf{return} $\text{campaign\_profile}, \text{NoRulesGenerated}$
    \ENDIF
    \end{algorithmic}
\end{algorithm}

The system's core strength is its adaptive Pivotal Analysis.
A central Supervisor Agent orchestrates the investigation by dynamically tasking Specialized Agents based on an evolving Knowledge Base (KB).
At each step, the Supervisor's LLM executes the \texttt{DecideNextAction(KB, History)} function (Algorithm~\ref{alg:supervisor}) to assess the collected evidence and determine the most promising pivot.
This approach is fundamentally different from static analysis pipelines.
This multi-agent design provides key advantages, including \textit{modularity}, potential \textit{scalability}, and \textit{separation of concerns}, allowing the Supervisor to focus on strategy while Specialized Agents gather data.

At the core of the loop, the LLM assesses the entire Knowledge Base to identify the most promising investigative leads based on a set of triggers.
These include: (1) \textit{Immediate Triggers}, such as significant cloaking behavior (e.g., a newly registered domain redirecting to an unrelated major site like \texttt{google.com}); and (2) \textit{Infrastructure Triggers}, where accumulated evidence reveals suspicious hosting patterns, such as multiple unrelated domains on a non–Content Delivery Network (CDN) IP address.
Importantly, content inaccessibility or cloaking alone does not constitute a positive decision; it only prompts a pivot to infrastructure analysis.
The campaign classification requires correlated signals, such as shared IP ranges, TLS certificate reuse, or registrar patterns, before any rule is produced.
This mechanism is resilient to domain rotation and short-lived anonymization, because the Supervisor escalates pivots from domains to IP ranges, Autonomous System Numbers (ASN), and certificate issuers when lower-level features become volatile (see Table~\ref{tab:case_studies}).
This state-aware, trigger-based orchestration allows PhishLumos to allocate computation where it is most informative, adapting its focus to the tactics used by each campaign.

\subsection{Data Acquisition and Context Management}
PhishLumos employs a systematic process for data acquisition and context management to support its adaptive investigations.
External intelligence is sourced programmatically from threat intelligence Application Programming Interfaces (APIs), such as \texttt{urlscan.io}~\cite{urlscanio} and \texttt{VirusTotal}~\cite{virustotal}, which provide access to essential datasets such as historical website scans, DNS records, and TLS certificate attributes.
Specialized Agents dynamically retrieve this information using dedicated API interaction tools.

To operate effectively within the context length limitations of its LLMs, PhishLumos implements a formal context management strategy.
This is crucial for distilling large volumes of raw data into a concise format suitable for LLM processing.
The strategy includes several key functions: \textit{selective data extraction} to filter raw API data for essential fields; \textit{structured inter-agent communication} to standardize findings into easily parsable summaries; \textit{hierarchical summarization}, where each agent's LLM identifies key patterns from its tool outputs; and \textit{dynamic context maintenance} by the Supervisor to integrate new findings without context overflow.
This integrated strategy enables the LLM components to reason effectively over pertinent information extracted from extensive external data sources, allowing for deep, multi-faceted investigations.

\subsection{Campaign Analysis and Rule Generation}
The final stage of the PhishLumos pipeline consists of two critical, sequential phases performed by its two Synthesis Agents: the Campaign Analysis Agent and the Rule Generation Agent.
This process corresponds to the synthesis phase of the workflow (lines 25--32 in Algorithm~\ref{alg:supervisor}), which is triggered after the iterative investigation loop concludes.

\begin{table}[t]
    \centering
    \notsotiny
    \caption{Evidence indicators for campaign classification.}
    \label{tab:campaign_heuristics}
    \setlength{\tabcolsep}{3pt}
    {\renewcommand\arraystretch{0.75}
    \begin{tabular}{ll}
    \toprule
    \textbf{Campaign Type} & \textbf{Primary Evidence Indicators} \\
    \midrule
    CONFIRMED & Consistent brand detection or content hashes, linked by shared IP addresses.\\
    \midrule
    CLOAKED & Discrepancy between input and final URL, especially to a known legitimate site.\\
    \midrule
    REUSE & Shared IPs/Certificates with historically malicious domains (content may differ).\\
    \midrule
    UNAVAILABLE & Inaccessible content (e.g., 404), but infrastructure linked to a known campaign.\\
    \midrule
    UNCLEAR & Insufficient correlated evidence to make a confident classification.\\
    \bottomrule
    \end{tabular}
    }
\end{table}

\noindent\textbf{Campaign Characterization.}
The Campaign Analysis Agent performs a comprehensive synthesis of the final Knowledge Base.
It employs an LLM guided by a set of formal heuristics, outlined in Table~\ref{tab:campaign_heuristics}, to correlate technical artifacts and determine a campaign profile.
This involves identifying shared IP infrastructure, common domain registration patterns, links via TLS certificates, and temporal clustering of activities.
The agent then assigns a campaign classification and distills its defining features into a structured set of \texttt{KeyIndicators} for the subsequent rule generation phase.

\noindent\textbf{Rule Generation and Validation.}
The extracted \texttt{KeyIndicators} are passed to the Rule Generation Agent, which executes the multi-step procedure in Algorithm~\ref{alg:rule_generation}.
First, its LLM generates a diverse set of candidate rules ($R_{cand}$).
Second, to ensure responsible deployment, each candidate rule is empirically validated.
This step is critical to prevent overly broad rules that could have high false-positive rates.
Such rules could inadvertently block legitimate services, causing negative social impact.
The agent queries a threat intelligence API for a small data sample and prompts its LLM to assess the rule's precision and contextual relevance.
For example, it checks if newly found domains share naming patterns or hosting characteristics with the initial seeds.
The final output is a set of validated rules ($R_{final}$), delivered as actionable threat intelligence (e.g., structured search queries).
This intelligence is designed for direct integration into security ecosystems, such as updating blocklists for web browsers and email gateways, or providing evidence for infrastructure takedown requests to hosting providers.

\begin{algorithm}[t]
    \notsotiny
    \caption{Campaign Rule Generation and Validation}
    \label{alg:rule_generation}
    \begin{algorithmic}[1]
    \STATE \textbf{Input:} Campaign Profile $P$ (with KeyIndicators $I_k$)
    \STATE \textbf{Initialize:} Candidate Rules $R_{cand} \leftarrow \emptyset$, Final Rules $R_{final} \leftarrow \emptyset$
    
    \STATE \COMMENT{Generate diverse rule proposals}
    \STATE $prompts \leftarrow \text{LLM.CreateGenerationPrompts}(I_k)$
    \COMMENT{e.g., ``Generate infrastructure-based rules'', ``Generate cloaking-based rules''}
    \STATE $R_{cand} \leftarrow \text{LLM.GenerateRulesFromPrompts}(prompts)$
    
    \STATE \COMMENT{Evaluate each candidate rule empirically}
    \FORALL{$r \in R_{cand}$}
        \STATE $query \leftarrow \text{FormatAsApiQuery}(r)$
        \STATE \COMMENT{Validate with a small data sample}
        \STATE $sample\_results \leftarrow \text{ThreatIntelAPI.search}(query)$
        
        \IF{$sample\_results$ is not empty}
            \STATE \COMMENT{Assess precision and coverage from sample}
            \STATE $precision, coverage \leftarrow \text{LLM.AssessSample}(sample\_results, P)$
            \STATE \COMMENT{Assess sample's relevance to campaign profile}
            \STATE $r.\text{score} \leftarrow \text{CalculateScore}(precision, coverage)$
        \ELSE
            \STATE $r.\text{score} \leftarrow 0$
        \ENDIF
    \ENDFOR
    
    \STATE \COMMENT{Select and recommend the best rules}
    \STATE $R_{final} \leftarrow \text{SelectBestRulesByScore}(R_{cand})$
    \STATE \textbf{return} $R_{final}$
    \end{algorithmic}
\end{algorithm}

\section{Experiments}
This section experimentally evaluates PhishLumos.
We first describe the experimental setup, including the datasets and implementation details.
Then, we present the results to demonstrate the system's effectiveness in (1) proactively mitigating entire phishing campaigns and its resulting social impact, (2) handling content-inaccessible scenarios where traditional methods fail, (3) generating practical intelligence, and (4) justifying its multi-agent architecture.

\subsection{Experimental Setup}

\noindent\textbf{Datasets.}
To evaluate PhishLumos, we constructed two primary datasets: a ground truth collection of phishing URLs and a set of benign URLs for false positive analysis.
Our ground truth phishing dataset was built using public data from the Japan Computer Emergency Response Team (JPCERT/CC)~\cite{JPCERT-feed}, spanning from January 2023 to March 2025.
JPCERT/CC is Japan's national Cyber Security Incident Response Team (CSIRT), and its data is highly reliable as human analysts manually verify each reported phishing URL.
For each URL, JPCERT/CC publishes the verification date and the targeted brand confirmed by a human analyst; we treat these fields as ground-truth labels.
We identified and validated 103 distinct campaigns, comprising 6,020 unique URLs.
For false positive analysis, we used a benign dataset of 1,000 safe URLs from the ChatPhishDetector study~\cite{DBLP:journals/access/KoideNC24}, including legitimate corporate and top-ranking websites~\cite{DBLP:journals/access/ChibaNK25}.

\noindent\textbf{The Challenge of Content-Inaccessible Scenarios.}
A critical challenge in modern phishing detection is the prevalence of evasive tactics.
Our analysis of the ground truth URLs revealed that automated web analysis APIs often fail to retrieve direct phishing content.
Cloaking or access errors affected 77.0\% of URLs (82.5\% of campaigns), preventing direct content retrieval.
This rate reflects failures of automated scanners at crawl time; it does not imply that human analysts could not observe phishing content when JPCERT/CC verified the URLs.
In PhishLumos, content unavailability is only a trigger to pivot to infrastructure analysis; the system does not assign a positive label on this signal alone.
This prevalence of content-inaccessible scenarios underscores the limitations of content-based detectors and highlights the need for infrastructure-focused analysis like PhishLumos.
A breakdown reveals diverse evasive tactics: 38.8\% used cloaking or deceptive redirection, 37.6\% were inaccessible or actively blocked scanners (e.g., with \texttt{403}/\texttt{404} errors), and the remaining 23.5\% presented ambiguous content with no clear malicious indicators.
The problem is escalating, with the rate of these content-inaccessible scenarios increasing from 48.7\% in 2023 to 95.4\% in 2024, and reaching 100\% for the campaigns observed in early 2025.

\noindent\textbf{Baselines.}
We compared PhishLumos with state-of-the-art content-based detectors: ChatPhishDetector~\cite{DBLP:journals/access/KoideNC24}, Phishpedia~\cite{DBLP:conf/uss/LinLDNCLSZD21}, and VisualPhishNet~\cite{DBLP:conf/ccs/AbdelnabiKF20}.
These systems primarily rely on accessible webpage content (text, images, logos) to classify individual URLs.
For PhishLumos, we treat the generation of a valid detection rule for a campaign as a positive classification for the initial URL that triggered the investigation.

\noindent\textbf{Implementation and Reproducibility.}
PhishLumos is implemented in Python 3.11.12, using the LangGraph library (v0.4.5) to manage the multi-agent workflow.
All agents use Azure OpenAI's \texttt{gpt-4.1-2025-04-14} model.
To ensure deterministic outputs and maximize reproducibility, key hyperparameters were set to fixed values.
We set a \texttt{temperature} of 0 and a \texttt{seed} of 42 for this purpose.
Given the deterministic nature of this setup, each experiment was conducted once per input URL.
For external data gathering, the agents interfaced with the public API of \texttt{urlscan.io}~\cite{urlscanio}.
All experiments were run on a MacBook Air with an Apple M2 CPU and 24GB of RAM, running macOS Sequoia 15.5.
To facilitate reproducibility, our research assets---source code, full dataset, agent prompts, and execution logs---are available upon request.

\noindent\textbf{Operational model and scalability.}
In deployment, we invoke the LLM agents only for a small set of seed URLs (e.g., Security Operations Center (SOC) triage).
The generated rules compile to lightweight predicates---regular expressions over domains and paths, IP/ASN filters, and TLS issuer fingerprints---that mail gateways and web proxies can evaluate at line rate.
This design decouples offline, higher-cost rule synthesis from online, low-cost matching; production systems apply only the rules, not the LLM, to traffic streams.

\begin{table}[t]
    \centering
    \notsotiny
    \caption{Overall performance of PhishLumos.}
    \label{tab:overall_performance}
    {\renewcommand\arraystretch{0.75}
    \begin{tabular}{lrr}
    \toprule
    \textbf{Metric} & \textbf{Mean} & \textbf{Median} \\
    \midrule
    Campaign Coverage (\%) & 93.0 & 100.0 \\
    New URLs Found (per campaign) & 751.4 & 297.0 \\
    \midrule
    Detection Lead Time (h) & 529.1 & 192.8 \\
    \midrule
    Execution Time (s) / URL & 63.9 & 60.4 \\
    Cost (\$) / URL & 0.25 & 0.23 \\
    \bottomrule
    \end{tabular}
    }
\end{table}

\begin{table}[t]
    \centering
    \notsotiny
    \caption{Performance comparison in different scenarios.}
    \label{tab:low_info_comp}
    \setlength{\tabcolsep}{5pt}
    {\renewcommand\arraystretch{0.75}
    \begin{tabular}{llrrrr}
    \toprule
    \textbf{Scenario} & \textbf{Method} & \textbf{Precision} & \textbf{Recall} & \textbf{F1-Score} & \textbf{False Positive Rate} \\
    \midrule
    \textbf{Content-} & PhishLumos & \textbf{0.947} & \textbf{1.000} & \textbf{0.973} & \textbf{0.001} \\
    \textbf{Accessible} & ChatPhishDetector & 0.621 & 1.000 & 0.766 & 0.011 \\
    (18 campaigns) & Phishpedia & 0.882 & 0.833 & 0.857 & 0.002 \\
    & VisualPhishNet & 0.033 & 0.667 & 0.064 & 0.347 \\
    \midrule
    \textbf{Content-} & PhishLumos & \textbf{0.988} & \textbf{1.000} & \textbf{0.994} & \textbf{0.001} \\
    \textbf{Inaccessible} & ChatPhishDetector & 0.725 & 0.341 & 0.464 & 0.011 \\
    (85 campaigns) & Phishpedia & 0.333 & 0.012 & 0.023 & 0.002 \\
    & VisualPhishNet & 0.000 & 0.000 & 0.000 & 0.347 \\
    \bottomrule
    \end{tabular}
    }
\end{table}

\begin{table*}[t]
    \centering
    \notsotiny
    \caption{Case studies demonstrating PhishLumos's reasoning. Note: Brand names and technical indicators are anonymized for security.  \textit{Initial} refers to the properties of the input URL, while \textit{Final} refers to the page attributes after any redirects.}
    \label{tab:case_studies}
    {\renewcommand\arraystretch{0.75}
    \begin{tabular}{lll}
    \toprule
     \textbf{Case Study} & \textbf{Generated Rule} & \textbf{Insight: Why this rule is intelligent and practical} \\
    \midrule
     1. Zero initial scan data& Initial Domain: \texttt{*.co.jp.*.test} AND & Pivoted from a dead-end URL to its domain, spotting a suspicious domain pattern.\\
     (dead-end URL) &  Date: \texttt{[2023-09-02 TO 2024-02-27]} &  This uncovered a regional campaign, finding 405 new URLs.\\
    \midrule
     2. Deceptive cloaking& Initial URL: \texttt{*/[unique\_string]*} AND & Identified a non-obvious URL path string as a unique campaign fingerprint, \\
     (redirect to benign site) & Date: \texttt{[2024-11-26 TO 2024-12-24]} & ignoring the misleading redirect. Discovered 540 new URLs.\\
    \midrule
     3. Advanced cloaking& Initial Domain: \texttt{*.suspicious.test} AND Final Domain: \texttt{official.example} & Defeated cloaking by flagging the mismatch between the suspicious initial and \\
     (redirect to official site) & AND Date: \texttt{[2023-04-24 TO 2023-05-15]} & legitimate final domains. Found 281 new cloaked URLs.\\
    \midrule
     4. Scanner blocking& Final IP: \texttt{203.0.113.43} AND Final Status: \texttt{403} & Turned a `403 Forbidden' error into a detection rule, treating the evasion tactic\\
     (403 Forbidden error)  & AND Date: \texttt{[2024-03-30 TO 2024-07-26]} &  itself as a cloaking signal. Identified 65 new URLs.\\
    \midrule
     5. IP hopping to& Final ASN: \texttt{AS64500} AND Final Domain: \texttt{*.test} & Created a resilient rule by targeting the stable ASN instead of transient IPs, \\
    evade blocklists. &  AND Date: \texttt{[2023-06-14 TO 2024-01-01]} & moving up the network stack. Preemptively found 4,118 URLs. \\
    \bottomrule
    \end{tabular}
    }
\end{table*}

\begin{table}[t]
    \centering
    \notsotiny
    \caption{Ablation study results.}
    \label{tab:ablation}
    {\renewcommand\arraystretch{0.75}
    \begin{tabular}{lrr}
    \toprule
    \textbf{System Configuration} & \textbf{Mean Coverage} & \textbf{Median Coverage} \\
    \midrule
    PhishLumos (Full System) & \textbf{93.0\%} & \textbf{100.0\%} \\
    w/o Pivotal Analysis & 56.1\% & 84.8\% \\
    w/o Domain Analysis & 62.4\% & 90.0\% \\
    w/o IP Analysis & 56.6\% & 84.8\% \\
    w/o Certificate Analysis & 60.2\% & 87.7\% \\
    w/o Campaign Analysis & 49.0\% & 48.1\% \\
    w/o Rule Generation & 26.5\% & 2.2\% \\
    \bottomrule
    \end{tabular}
    }
\end{table}

\subsection{Proactive Mitigation and Social Impact}
We first evaluated PhishLumos's primary goal: its ability to uncover and mitigate entire phishing campaigns from a single URL.
Table~\ref{tab:overall_performance} summarizes the key results.

PhishLumos demonstrated excellent performance, achieving a median campaign coverage of 100\%.
This indicates that the generated rules were effective at identifying nearly all known URLs within a given campaign.
The system showed powerful proactive detection capabilities.
It discovered 77,391 URLs not present in our original ground truth data.
We evaluated these new URLs using \texttt{VirusTotal}~\cite{virustotal}, which uses results from 90 security engines.
Of these URLs, 62,321 (80.5\%) were later flagged as malicious by at least one engine.
This high confirmation rate is conservative, as our method finds cloaked sites before they are blocklisted.
The fact that nearly 20\% of these newly discovered URLs still evaded this extensive panel of scanners underscores the sophistication of the threats PhishLumos uncovers.
To ensure practical safety, a parallel check against top-ranking websites~\cite{DBLP:conf/ndss/PochatGTKJ19} confirmed zero false positives, underscoring the precision of our method.
On average, this translates to discovering 751.4 new malicious URLs per campaign.

The system's primary social impact is its proactive nature.
We measured the \textit{detection lead time} as the interval between PhishLumos generating a rule and the URL's initial verification by JPCERT/CC.
This timestamp marks the point when a human analyst confirms the site as malicious, a critical first step in the official mitigation and takedown process.
The system identified campaigns a median of 192.8 hours (approximately 8 days) before this expert verification.
This significant lead time enables practical, preemptive defense.
For example, the generated intelligence allows browser vendors and security firms to deploy blocking rules network-wide, effectively shutting down attack vectors before they can cause widespread financial and emotional harm to vulnerable populations.
The analysis is efficient, taking a median of 60.4 seconds per URL.

\subsection{Performance on Evasive Targets}
The system's primary advantage is its effectiveness in content-inaccessible scenarios, where conventional content-based detectors struggle.
We evaluated PhishLumos against baselines by dividing our dataset based on whether an automated scanner could access the final phishing content.

As detailed in Table~\ref{tab:low_info_comp}, the performance of all content-based methods deteriorates significantly in content-inaccessible environments.
Vision-based detectors like Phishpedia and VisualPhishNet become ineffective, as their F1-Scores fall to 0.023 and 0.000, respectively.
This is due to the absence of visual cues like logos or site layouts.
Even the state-of-the-art LLM-based ChatPhishDetector struggles, with its recall dropping to 0.341 because it lacks sufficient textual content to analyze.

In sharp contrast, PhishLumos maintains nearly perfect performance, achieving an F1-Score of 0.994 with 1.000 recall.
This resilience comes from its core design: instead of relying on ephemeral content, it analyzes stable infrastructure patterns.
This approach allows it to reliably identify threats even when a site's malicious content is cloaked or the page itself is made inaccessible.
Furthermore, even when content is available, PhishLumos achieves the highest F1-Score (0.973), outperforming all baselines while maintaining a minimal False Positive Rate of 0.001 across all scenarios.
To validate these performance gains, we conducted McNemar's tests, which confirmed that PhishLumos achieved a statistically significant improvement over all baselines in both evaluation scenarios ($p<0.001$ for all comparisons).
This highlights a fundamental advantage of shifting the detection focus from volatile content to persistent infrastructure footprints.

\subsection{Qualitative Analysis with Case Studies}
Table~\ref{tab:case_studies} presents five case studies that highlight the system's ability to generate high-quality, actionable detection rules.
These cases show the system's ability to pivot from zero-information starts (Case 1), discover non-obvious URL path fingerprints (Case 2), identify sophisticated cloaking (Case 3), generalize from server responses (Case 4), and create robust rules by targeting the Autonomous System Number (ASN) level (Case 5).
The Rule Generation Agent played a crucial role by automatically discarding overly broad or transient rules, ensuring the final output is both precise and durable.

\subsection{Ablation Study of System Components}
Finally, we conducted an ablation study to validate our multi-agent design.
The results in Table~\ref{tab:ablation} confirm that every component is critical for the system's performance.

The most significant performance degradation was observed in the \textit{w/o Rule Generation Agent} configuration, where median coverage dropped to 2.2\%.
This experiment does not mean rule generation was skipped entirely.
Instead, it tested a naive approach where the system generates only a single candidate rule without the subsequent process of generating multiple hypotheses (7--10 candidates) and validating them.
This result highlights that proposing and evaluating diverse candidate rules is the most critical factor for creating effective and comprehensive campaign definitions.

Furthermore, disabling the Pivotal Analysis capability, the core of our approach, also caused a substantial drop in mean coverage from 93.0\% to 56.1\%.
This confirms that investigating peripheral infrastructure is essential.
The removal of any other agent also led to a significant decline in performance, justifying our multi-agent architecture.

\section{Conclusion}
Phishing poses a severe threat to society, disproportionately harming vulnerable individuals.
We developed PhishLumos, a system to proactively mitigate attack campaigns before they cause widespread damage.
Instead of relying on often-inaccessible content, PhishLumos analyzes stable infrastructure patterns to uncover entire campaigns with high precision.
Our work provides a crucial detection lead time, creating a vital window to protect users.
This research demonstrates a practical shift from reactive defense to proactive mitigation, making essential online services safer for everyone.

\bibliographystyle{IEEEtran}
\bibliography{main}

\end{document}